# Data-driven State of Risk Prediction and Mitigation in Support of the Net-zero Carbon Electric Grid

Mladen Kezunovic, Rashid Baembitov, Mohammad Khoshjahan

Department of Electrical and Computer Engineering

Texas A&M University

College Station, Texas, USA 77843

m-kezunovic@tamu.edu, bae_rashid@tamu.edu, mohammad.kh@tamu.edu

*Abstract*—An approach for reaching the net-zero carbon electricity grid is to intensify the deployment of distributed renewable generation resources such as photovoltaic (PV) solar and wind generation, complemented with stationary and mobile (electric vehicle) battery energy storage systems (BESS). This paper assumes a scenario where the PV renewable generation and BESS are integrated into a nano-Grid (n-Grid), a prosumer-owned virtual power plant installed at the residential or commercial sites. To be profitable, this distributed energy resource (DER) needs to be managed effectively to support its own load and the wholesale and retail market services. Our paper introduces a risk-based, data-driven approach focused on predicting the State of Risk (SoR) of the utility grid outages. The SoR prediction enables the development of optimal mitigation strategies aiming at reducing the impact of the grid outages by harvesting the n-Grid flexibility. Several data analytics tools to assist the n-Grid operators and aggregators for n-Grid participation in the wholesale market ancillary service products are introduced, and some preliminary implementation results are demonstrated.

## Nomenclature

*Abbreviations*

| | |
|---|---|
| PV | Photovoltaic |
| BESS | Battery energy storage systems |
| n-Grid | nano-Grid |
| DER | Distributed energy resource |
| SoR | State of Risk |
| ML | Machine Learning |
| DM | Data-driven model |
| ASP | Ancillary service products |
| WEM | Wholesale electricity market |
| FERC | Federal Energy Regulatory Commission |
| EVCS | Electric Vehicle Battery and Charging Station |
| DSO | Distribution system operator |
| CEL | Controllable Electric Load |
| HVAC | Heating, ventilation, and air conditioning |
| GIS | Geographical Information System |
| ASOS | Automated Surface Observing System |
| NOAA | National Oceanic and Atmospheric Administration |
| NAIP | National Agriculture Imagery Program |
| ROC AUC | Area Under the Receiver Operating Characteristic Curve |
| PRC AUC | Area Under the Precision-Recall Characteristic Curve |
| FM | Final Metric |
| ISO | Independent System Operators |
| HEMS | Home energy management system |
| SR | Spinning reserve |
| NSR | Non-spinning reserve |
| AGC | Automatic generation control |
| ENS | Energy not served |
| RU | Ramp up |
| RD | Ramp down |

## Introduction

The world political powers are slowly converging on a consensus to reduce carbon emissions resulting from the use of fossil fuel derivatives for electricity generation and vehicle transportation. While the renewable generation utilizing solar radiation and wind force look attractive, they are inherently highly variable as the weather constantly changes and lately shows more extreme behavior than before. The use of BESS is frequently mentioned as the approach to mitigate the variability and introduce more certainty in the grid operation. The nano-

This material is based upon work supported by the U.S. Department of Energy under Award Number DE-IA0000025. The views and opinions of the authors expressed herein do not necessarily state those of the United States Government or any agency thereof.

Grid (n-Grid) architecture shown in Fig. 1 can be used to implement DERs that will facilitate net-zero carbon grids.

We expand the mitigation of uncertainties due to weather conditions that cause grid outages by harnessing the flexibility of customer-owned nano-Grid (n-Grid) [1]. This requires formulation of the risk-based approach enabling the selection of optimal mitigation strategies utilizing n-Grid resources to reduce and even avoid outage impact. Machine Learning (ML) combined with data-driven models (DM) and advanced optimization techniques are used to predict the State of Risk (SoR) of outages allowing sufficient time to deploy mitigation strategies.

Since the outages may occur either due to sudden imbalances between load and generation or due to the faults associated with the impacts of the inclement weather, the focus is on two approaches: a) how to mitigate outage impact at the local utility feeder using n-Grid resources, and b) how to aggregate such distributed resources to manage the adverse power system conditions through the participation of n-Grids in the ancillary service products (ASP) at the wholesale electricity market (WEM). The latter is receiving further attention after the FERC issued Order 2222, which requires market operators to enable the participation of DER aggregators in ASP markets [2].

Prior work on managing the uncertainty of renewables is very comprehensive and elaborate [3-7]. The specific use of prosumers or virtual power plants for this purpose is also quite diverse [8-12]. Combining such approaches with the risk-based analysis is studied in [13]. The use of DM and ML to predict the state of risk is relatively recent [14, 15]. The specific utilization of n-Grids to achieve their resource flexibility to help the n-Grid owner meet its load needs during faults and to support WEM and APS during utility grid emergencies is discussed in several most recent papers [16-18]. The SoR prediction, combined with the utilization of DM, has been instrumental in developing the operator and aggregator decision-making tools to manage outage contingencies [19-23].

Our previous works covered the advantages that optimal management of n-Grids brings about to the utility grid [1], [24-26]. Hence, in this paper, we concentrate on the advantages of n-Grid participation in the WEM through aggregation. Our contribution is the inclusion of the DM-based weather-induced utility outage SoR predictions in the aggregator and n-Grid owner resource management tools, which provide users with the means to mitigate outage impacts. More specifically, we link a predictive SoR outcome to an optimization approach for scheduling n-Grid resource services in the most profitable way for the owner and other stakeholders. We illustrate this approach using data from actual utility systems, and we present the results utilizing real-time data. The benefits of the approach are not solely limited to the n-Grid owner, aggregator, and utility. Also, as an intensive user of the electric grid power delivery, the society at large experiences a decrease in the costs incurred by the impact of outages and system instabilities through resiliency and reliability enhancements. The risk of cascading outages is also reduced, preventing catastrophic damage to the economy and consumer well-being [27, 28]. As United States Government has pledged to reach net-zero greenhouse gas emissions by 2050 [29], we are not considering non-environmentally friendly resources, such as backup diesel generators, for reliability improvement.

The paper first introduces the regulatory framework of aggregators in WEM in Section II. The description of the n-Grid architecture and its flexibility is given in Section III. We then introduce the risk-based analysis approach for outage SoR prediction in Section III. The n-Grid utilization of the SoR prediction to mitigate outage impacts is discussed in Section IV. Some preliminary results are shown in Section V. Conclusions and references are given at the end.

I. REGULATORY FRAMEWORK FOR PARTICIPATION OF AGGREGATORS IN THE WEM

The federal energy regulatory commission (FERC) Order 2222 mandates the US ISOs to prepare for the participation of DER aggregators in the wholesale energy and ASP markets [2]. This Order is a milestone in the DER aggregation development in the US by providing a possible profitability opportunity for aggregators and n-Grid owners. At the same time, the ISOs can benefit from extra sources of flexibility to deal with the intermittencies stemming from the high penetration of renewable energy sources. The impacts of this Order on aggregators and n-Grids are discussed below.

*A. Impacts of the FERC Order on Aggregators*

Based on this Order, aggregators can procure ASPs for and trade energy in the WEM. The n-Grid resources, namely, EV availability and state-of-charge, deferrable and thermal loads, are naturally variable. To avoid penalties associated with the inability to deliver the rewarded ASPs, the aggregator must perform proper forecasts and optimization algorithms to ensure the availability of DERs. Aggregators can benefit from the SoR predictions to obtain the risks of the unavailability of n-Grid resources and act accordingly.

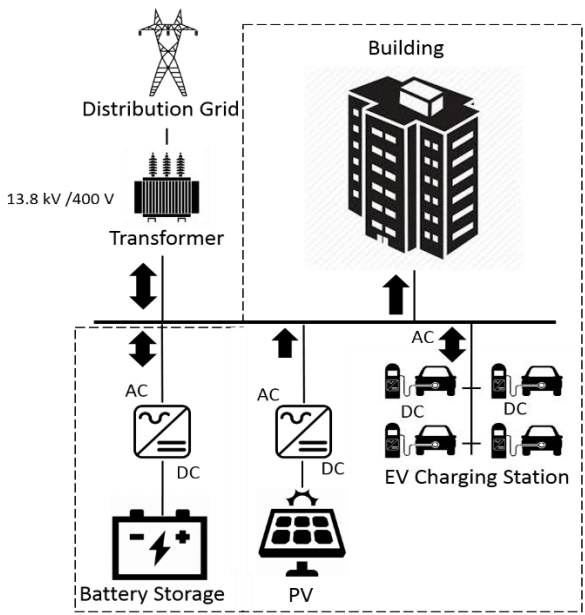

Figure 1. The n-Grid architecture [1].

It should be mentioned that according to the Order, the aggregator does not need to be concerned with the distribution grid constraints, such as voltage and line capacity limits, since it does not need the permission of the DSO to participate in the WEM [2].

### B. Impacts of the FERC Order on n-Grid Owners

The n-Grid owners can profit from this opportunity by enabling aggregators to control their resources. The more flexibility they provide, the higher their profit may be. They can also benefit from SoR predictions by preparing for the times with high risks of outages and planning on having less unsupplied load during the outages. They store energy to supply their loads during outages and offer ASPs when they are connected to the grid. While the distribution utility or DSO cannot mandate aggregators or n-Grid owners to behave in a certain way [2], they may offer competitive incentives to aggregators or n-Grid owners to provide flexibility to the distribution grid, which they, in turn, can compare such benefits to the benefits of participating in the WEM.

## II. N-GRID RESOURCE FLEXIBILITY

We define the term nano-Grid (n-Grid) as a form of Prosumer or Virtual Power Plant, which is reduced to a commercial or residential building containing rooftop PV panels, a fixed BESS, EV mobile battery storage, and charging stations, and controllable electric loads [1]. If properly managed, the n-Grid is capable of serving its own load during contingencies as well as offering support for the grid. In the architecture of the n-Grid shown in Fig. 1, due to their high ramp rates, the n-Grid energy storage and generation resources can improve the system's reliability and resilience by procuring ASPs such as spinning reserve and frequency regulation for the WEM. In addition, distribution utility Operator (DSO) can benefit from optimized asset utilization and management practices in various grid support services such as voltage regulation and transformer wear and tear mitigation [25, 26]. Also, having the storage capability, an n-Grid owner can participate in energy arbitrage, that is, storing the energy when the prices are low and injecting power to the grid when the prices are high [24]. The flexibility of each of the n-Grid resources is described next:

*1) Battery Energy Storage System (BESS)*: The BESS has the highest value among the resources of n-Grid since it is always available (except the times of failure), and the uncertainty in its energy generation/storage due to battery degradation is manageable. It can change its state of operation and achieve its maximum power output in seconds. This feature can be harnessed to provide a backup power source for the n-Grid load and utility system. The BESS can also be used to gain profit from energy arbitrage. It is also a desirable solution for mitigating the impacts of uncertain PV generation in n-Grids and providing support for renewable energy generation resources in the utility grid [30, 31].

*2) Electric Vehicle Battery and Charging Station (EVCS)*: The EV battery, when connected to the EVCS, brings about the same advantages as the BESS described earlier, except it is tied to its mobility. The size of the EV battery is usually larger than the local BESS, which brings even higher flexibility to the system. However, the flexibility of this resource is impeded by the uncertainties in the EV spatiotemporal availability and initial stored energy. Thus, planning on utilizing the EV battery flexibility is challenging [32, 33].

*3) Controllable Electric Loads (CELs):* The CELs are another source of flexibility provided by n-Grid. CELs are divided into adjustable loads and deferrable loads. The building's heating, ventilation, and air conditioning (HVAC) can be referred to as an adjustable load. The building occupants set their temperature comfort range by the thermostat. The building temperature and, consequently, HVAC load can be adjusted in the preset comfort range to enhance flexibility. The deferrable loads such as laundry and dishwashing machines can be postponed as needed to later times. In summary, the building occupants can provide input to an algorithm that selects time intervals to run the n-Grid load based on the requirements of the bulk power grid. Such loads have uncertainties raised from the building occupancy times, temperature comfort range, type of deferrable load, and the time horizon for the task to be done [34-36].

## III. STATE OF RISK PREDICTION

The incorporation of DERs and renewables have been shown to have a positive emission reduction effect on the power system. Thus, assessing their ability to support normal operation and procurement of ASPs creates an additional incentive for their extensive adoption, leading to a further decrease in traditional carbon-intensive generation and a reduction in $CO_2$ emissions [37-40]. State of Risk (SoR) prediction can provide the necessary quantitative measure of the availability of the utility grid and n-Grid resources. Aggregators of n-Grids, n-Grid owners, utilities as well as consumers need decision-making tools for anticipating the risk impacts, which offers an opportunity for decreasing the level of uncertainty of the impacts through mitigation. In such a way, the volatile renewables turn into more stable resources with projected levels of commitment for continuous grid service delivery. At the same time, the net-zero carbon operation of the grid becomes feasible. Prediction of faults in the distribution systems for different time horizons using SoR levels can help determine the connectivity of n-Grids to the system at each moment of time, making it possible for aggregators to tune their market strategies. At a transmission level, the SoR for distinct geographic locations helps plan for congestions in the system, which allows for creating monetary incentives for WEM participants in different regions. The SoR prediction can also be used to assess the outage duration and, consequently, outage impacts. The availability and planned workload of dispatch crews, as well as their geographical location in relation to a fault location, are accounted for when determining the mitigation measures. Thus, the n-Grids themselves can adjust their power consumption to have enough energy for the whole outage duration. At the same time, aggregators receive real-time estimates of when the n-Grid is going to be connected back to the system and what its resource availability is going to be.

The interrelation of external weather and other environmental conditions, equipment and generation asset susceptibility to a fault caused by such conditions, SoR predictions, and risk management and mitigation strategies are illustrated in Fig. 2. Following the figure clockwise, the presence of severe weather conditions may cause a severe *hazard* to grid operations. Once such hazards are predicted, the *vulnerability* of the grid's equipment and energy resource assets can be predicted, leading to the prediction of the SoR. The SoR grid maps overlaid with Geographical Information Systems (GIS) enable timely assessment of SoR impacts. Then, an optimization objective can be defined to identify corrective actions for the grid operations management. The suggested actions, in turn, can guide market participants (aggregators) to plan *mitigation* strategies utilizing information about future n-Grid contingency scenarios. Given the historical analysis of market participants' behavior with the addition of spatiotemporal SoR maps, long-term planning, operations, and operations planning, maintenance, and real-time energy resource allocation can be optimized to reduce or even avoid outage impact rendering the electric energy delivery more resilient and reliable. Having described the overall approach of risk-enhanced power grid operation, we now focus on deeper aspects of mitigation strategies deployed by the n-Grid aggregator in WEM.

### A. SoR prediction implementation

The State of Risk (SoR) of any part of the power system at an arbitrary moment in time reflects the probability of its failure due to selected reasons, given the existing environmental conditions that directly or indirectly affect it. The SoR can be predicted ahead of time by the coalescence of GIS, DM, and ML algorithms [14, 15, 41]. By the spatiotemporal tacking of the SoR, the utility grid operators can optimize their resources, and n-Grids can execute a set of mitigation actions (subject to constraints) to either lower the SoR to their own load or adjust their strategy of participation in the WEM's ASPs. Such an approach results in increasing utility grid reliability and resilience for the consumer benefits, as well as maximizing profits of the market (aggregator) participants [13, 42-44].

The general approach to the prediction of SoR of various parts of the systems or an entire system shown in Fig. 2 leads to the detailed steps described in Fig. 3: a) Record of the historical performance of a utility infrastructure, which can be a single asset, a group of similar assets or a particular part of the system over a specific period, is obtained, b) Data reflecting various external conditions (weather, electrical parameters, periods of maintenance and replacements, disturbances, market conditions, etc.) are wrangled (identified, fetched, cleansed, preprocessed) into a suitable form, c) Spatiotemporal correlation of all datasets is utilized for training ML model(s) to predict SoR, d) The SoR maps are used to correlate existing conditions to the performance of the asset at a specific moment in time and particular geographic location [41], and e) The resulting SoR is utilized to devise different mitigation measures by stakeholders [13]. This paper only deals with optimization strategies of n-Gird aggregator in WEM as a mitigation measure. The utility implementations of mitigation strategies with respect to predicted SoR is a subject of future research.

We illustrate how, with sufficient collection of data, n-Grid owners can estimate their own levels of consumption and generation, which would help them to define their commitment to participate in WEM through an aggregator. The assumption is that the more n-Gird commits to providing extra services to the grid, the more it is getting paid. At the same time, to avoid compromising the comfort level of its own users, their risk tolerance needs to be considered when reaching a balance between profitability and comfort importance.

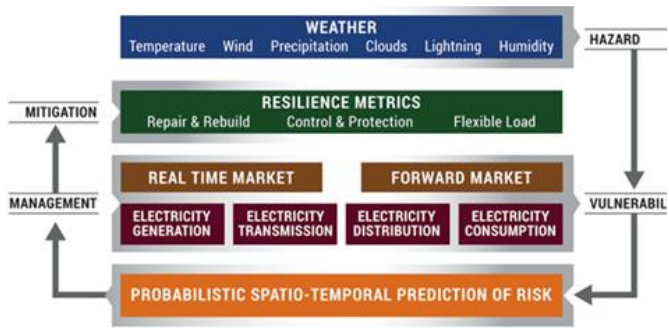

Figure 2. SoR in grid operations

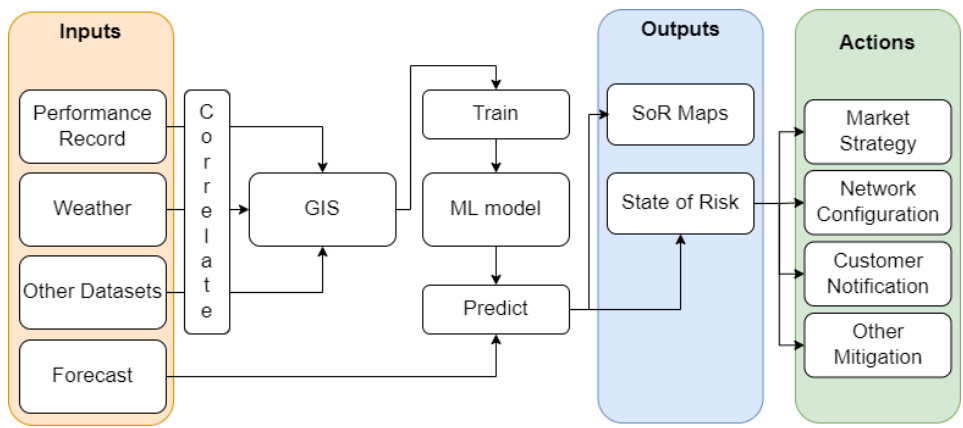

Figure 3. SoR framework diagram.

It is assumed that the utility provides the historical outage dataset, which has information about when, where, and why an outage has occurred in the system. The utility also outlines the location of the distribution feeders in the form of Geodatabase. It is used to associate the location of outages with actual equipment in the system. Since one of the main underlying reasons for outages is the weather, ASOS historical weather dataset provided by NOAA is used to correlate weather parameters to the periods of outages and normal operations [45, 46]. Additional datasets are incorporated into the ML model inputs: lighting dataset, NAIP imagery dataset, soil, and vegetation dataset. The datasets are wrangled to a suitable form before performing the spatiotemporal correlation. The 4-edge graph is formed and afterward transduced into several input features (dimensions) to improve the model's accuracy. Graph incorporation allows the capture of the spatial dependencies between different feeders [47].

The training dataset is constructed and fed into the ML algorithm. We used CatBoost, a type of gradient boosting algorithm(s) based on decision trees [48-50], because it can seamlessly ingest categorical features and output the feature importance table. Part of the data is separated to validate and test the trained model.

The performance of the trained prediction model is assessed using a set of metrics: ROC AUC, F-1 score, and PRC AUC [51]. In essence, ROC AUC reflects how well the model can separate one class (faults) from another (no fault or normal operation). The advantage of ROC AUC is that it is threshold-invariant because it reflects the performance of the model's predictions independent from the selected classification threshold [52]. F-1 score is a harmonic mean of precision and Recall. Precision shows the ratio of correctly predicted faults (true positives) to the total predicted faults (sum of true positives and false positives). Recall (also referred to as sensitivity) is the percentage of correctly predicted faults to all existing faults. Precision and Recall could be thought of as measures for type 1 and type 2 errors of the model [53]. PRC AUC is closely related to the F-1 score and indicates the mean of precision and Recall for all recall thresholds, whereas F-1 reflects the mean for the middle threshold – 50%.

Hyperparameters of the algorithm are tuned at this point, rendering a better performance. If the resulting accuracy of the model is deemed unsatisfactory, the model is modified by incorporating additional datasets, changing the underlying algorithm, or both.

An example of performance metrics scores for an ML prediction model is provided in Table I [47]. The model was trained and evaluated on 3 years of real-life data from a utility. The final metric (FM) is a weighted average of all metrics, as shown in eq. (1), which is one of the ways to compare different competing models. The results suggest that the use of ML prediction models accounts for inclement weather impacts and allows for accurate predictions of SoR in the system under various conditions.

$$FM = 0.4 \cdot ROC\_AUC + 0.3 \cdot F1 + 0.3 \cdot PRC\_AUC \quad (1)$$

TABLE I. MODEL PERFORMANCE METRICS [47]

|  | ROC AUC | F1 Score | PRC AUC | FM |
|---|---|---|---|---|
| Prediction Model | 0.939 | 0.856 | 0.944 | 91.57 |

The SoR maps correlate the system topology presentations with the embedded SoR levels. The SoR maps combined with the new decision-making tools help utilities' personnel quickly assess the possible impact of the risk levels in the system, allowing them to account for and prepare mitigation measures for different upcoming system contingencies. The example of an SoR map with optimal graph overlay is shown in Fig. 4, where distinct colors represent various levels of the SoR.

IV. RISK PREDICTION ENHANCED GRID SUPPORT THROUGH N-GRIDS PARTICIPATION IN WHOLESALE MARKET

Large-scale integration of n-Grids with utility grid is a promising approach to achieving high penetration of renewable resources. The n-Grid aggregators can procure ancillary service products (ASPs) for the wholesale electricity market (WEM), provide grid support services for the distribution system, and participate in peer-to-peer energy trading schemes. The key to the versatility of n-Grid services is in the flexibility of its energy resources that can be optimized to reach different objectives of SoR impact reduction.

*A. N-Grid Aggregation*

Fig. 5 depicts the n-Grid as a relatively small agent with many sources of uncertainty, e.g., load, PV generation, EV's battery initial state of charge, EV arrival and departure times, building occupancy, etc. Since participation in WEM ASPs typically requires a minimum power capacity between 100 kW and 1 MW, depending on the market [54, 55], the aggregator needs to engage a large number of n-Grids to participate in the WEM or distribution grid ancillary services.

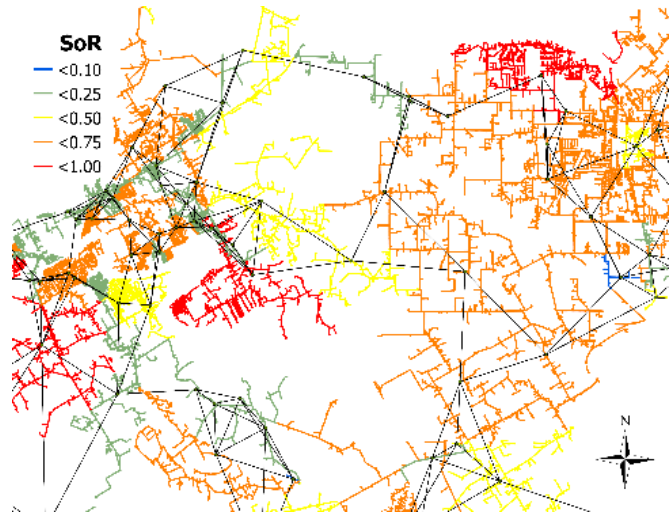

Figure 4. SoR Map.

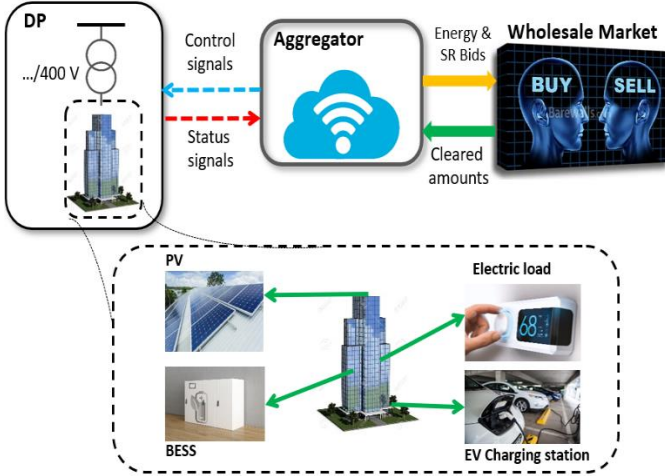

Figure 5. The aggregator/DP interaction [24].

As depicted in Fig. 5, the aggregator acts as the mediator between the n-Grids and WEM. Using historical data and probabilistic methods, the aggregator forecasts the n-Grid uncertainties and receives n-Grid predicted resource availability. It then runs optimization scenarios on SoR predictions using its computational resources to obtain an optimal market participation strategy to maximize its profitability. If the aggregator does not have the n-Grid SoR estimates, then it is forced into deciding about its market participation under a high risk of uncertainties. This potentially may result in higher risks to aggregator's bids, eventually leading to lower investment returns and profitability.

Next, the ISO clears the market based on the bids submitted by market participants and informs them about the cleared energy transactions and outcomes of ASP bids and their associated marginal prices. After the market clearance process, the aggregator stays committed to delivering the rewarded ASPs in real-time or otherwise faces financial penalties.

It is assumed that the aggregator can directly dispatch the n-Grid that signed up as the aggregation resources through the home energy management system (HEMS). In this regard, the aggregator sends the control signals to the n-Grids and receives the status signals in return. The control signal includes desired power output of all resources and state of operation for stationary and mobile BESS and EVCS. The n-Grid status signal includes the current metered power output of all resources and the current state of operation of its BESS and EVCS.

*B. N-Grid Participation in Ancillary Service Products*

Using the approach defined in the previous section, the n-Grid aggregation can offer multiple ASPs to the WEM. These ASPs include spinning reserve (SR), non-spinning reserve (NSR), frequency regulation, and flexiramp. The SR and NSR are procured for contingency response in the bulk power system. The resource must be able to deliver the rewarded capacity within 10 minutes. The resource must be synchronized with the grid for SR eligibility. However, for NSR, the resource must be able to synchronize with the grid and deliver the capacity in 10 minutes. The resources must be equipped with the automatic generation control (AGC) system for frequency regulation service procurement. The dispatch timeframe of this service is ~4-6 sec, depending on the market structure [54, 55].

The process of procuring SR from the n-Grid resources is shown in Fig. 6. The BESS can offer SR capacity in discharging mode up to the difference between maximum power output and current power. It can offer SR up to the current power output in charging mode. The SR capacity of the BESS is limited to its current state of charge. A similar requirement applies to the EV battery, except that EVs can offer SR capacity as long as they are plugged into the building charging station. Last but not least, the HVAC can offer SR up to its current power output. However, the reduction in power output must not violate the temperature range deemed acceptable by building occupants.

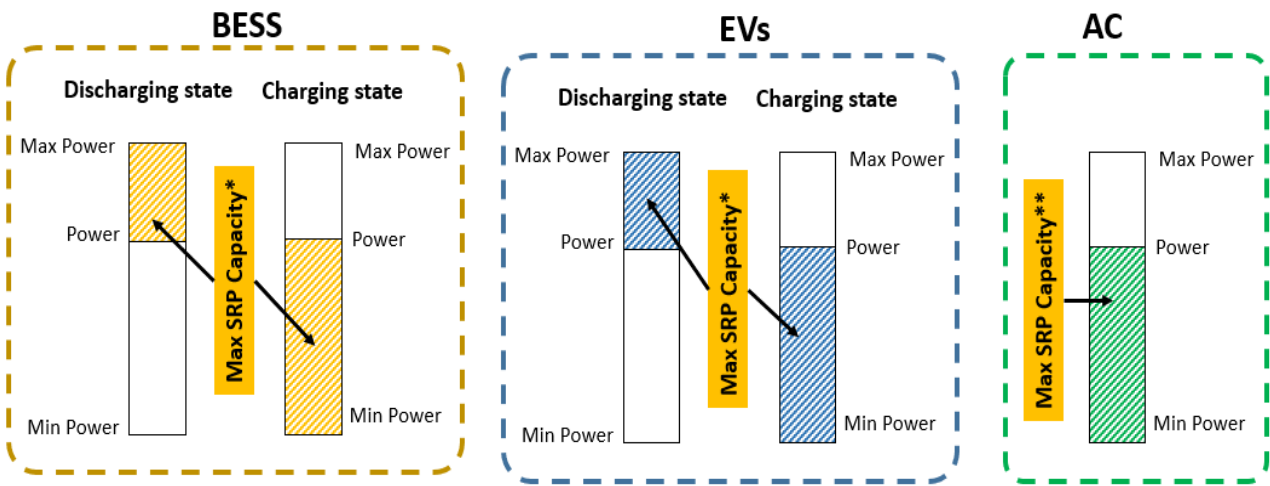

Figure 6. N-Grid resource employed for SR procurement [1].

\* The SR capacity offered by BESSs and EVs is limited to the current stored energy.

N-Grids can only deliver ASPs as long as they are connected to the grid. Hence, using SoR prediction results is a better n-Grid resource planning to improve their performance by optimal scheduling availability of their resources.

## V. Discussion of Use Cases and Results

In this section, we analyze the impacts of the weather-related outages on the n-Grid participation in the WEM. We consider the n-Grid aggregator participation in the WEM energy and spinning reserve markets, considering the outage probability of distribution feeders to which the n-Grids are connected. On this basis, the aggregator runs an optimization with the aim to maximize its profitability from energy trading and spinning reserve procurement in the day-ahead market. The optimization is subject to the technical constraints of n-Grid assets, including the minimum and maximum power and stored energy of BESSs and EVs, EV availability, minimum and maximum power of electric loads, etc. We study several scenarios (i) how the n-Grids in the faulted area can manage their load to minimize the expected energy not served (ENS) and spilled energy of the grid and (ii) the amount of flexibility (ramp capacity) the n-Grids connected to utility grid can offer.

For each scenario, we discuss how SoR prediction impacts the utilization of renewable energy resources toward the net-zero grid goals. The n-Grid flexibility is shown to enable the grid loads to ride through many contingencies despite or because of the deployment of large-scale renewables since the SoR prediction offers a time horizon outlook needed to define and implement mitigation measures to deal with the contingencies.

### A. Resource Management in Disconnected n-Grids

The Order of asset management for the disconnected n-Grids is given in the flowchart in Fig. 7. First, the deferrable load is delayed to later times, and the thermal demand is decreased to its minimum level set by building occupants. Now, if the net-load (load minus PV) is positive, the BESS is called to generate power. If still the net-load is not sufficiently supplied, the connected EVs are called to discharge power from their batteries. Any unsupplied demand is considered as the ENS of the n-Grid. The logic for this Order is that delaying the deferrable demand and reducing the HVAC demand may impact the n-Grid operation and occupants the most. The BESS and EVs' batteries are used next, but since the outage duration is uncertain, they may be needed later. The BESS is used before the EVs' battery since the EV owners may need the EV battery energy for driving. If the net-load is negative, the extra energy is first stored in the EVs' battery, and if any energy supply is left, it is stored in the BESS. If there is still extra energy supply left, it will be first used to meet the n-Grid thermal demand and then the deferrable demand. The extra energy is the spilled energy.

### B. N-Grid Performance with and without SoR Prediction

| *Algorithm: Asset management for disconnected n-Grids* | |
|---|---|
| | Delay the deferrable loads for later times |
| 1: | Decrease the HVAC load to its minimum level set by building occupants. |
| 2: | If remaining load ≥ PV: |
| 3: | Discharge BESS |
| 4: | If remaining load > PV + BESS: |
| 5: | Discharge EVs if connected |
| 6: | The extra load is the ENS |
| 7: | If remaining load < PV: |
| 8: | Charge EVs if connected |
| 9: | If remaining load < PV − EVs: |
| 10: | Charge BESS |
| 11: | If remaining load < PV − EVs − BESS: |
| 12: | Turn the AC thermostat back to normal |
| 13: | If remaining load < PV − EVs − BESS − AC: |
| 14: | Supply the deferrable load |
| 15: | The extra generation is the spilled energy |

Fig. 7 Asset management algorithm for disconnected n-Grids

In the following scenario, we consider that the n-Grids are connected to the laterals of 10 distribution feeders selected from the local utility in the Houston area. We assume that a fault in a feeder leads to the outage of the entire feeder. We assume 50 residential n-Grids are connected to each feeder, making a total of 500 n-Grids. Also, 750 EVs are connected to the n-Grids. The load and PV data of n-Grids are given in [56]. It is assumed that half of the n-Grids are equipped with the on-site BESS. The weather data for 23/02/2016 is considered for the case study. On this day, inclement weather was present in the location of the grid. Fig. 8 shows hourly predicted SoR levels for ten feeders (color-coded) calculated the day before. Such a time horizon allows planning for the day-ahead market. For the outage, the average repair time is set to 1 hr.

The total available ramp up (RU) and ramp down (RD) of the n-Grids are depicted in Fig. 9. Total RU and RD refer to the corresponding values if no outages occur, and the available RU and RD are the realized values due to outages. As can be seen, the total RU and RD are low from 7 am to 6 pm and are high at other times. The reason is that the building occupants tend to leave during the day and the EVs are primarily unavailable at the home location. Also, during harsh weather, the available RU and RD may drop significantly from the total values due to battery charge deterioration during weather extremes. This figure shows the significant capability of n-Grids in provisioning the ramp capacities to the grid with a maximum of approximately 7 MW from 500 residential n-Grids.

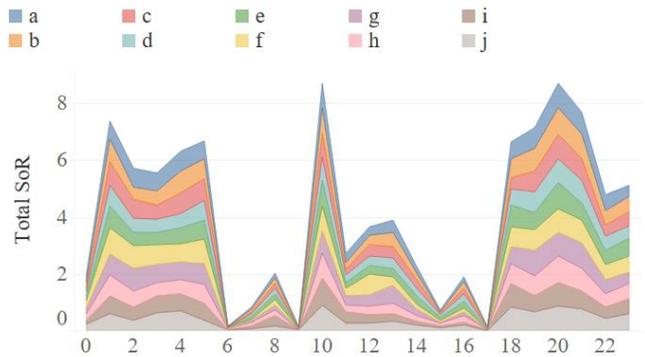

Figure 8. Hourly SoR Levels

The total ENS and spilled energy of the disconnected residential n-Grids are given in Fig. 10. By comparing the curves associated with the total load and load loss, it is observed that the n-Grids have a sufficient capability to supply their loads during outages. The total load loss in this case study was 0.01 MWh. Their performance in preventing PV generation spillage was not as significant. The reason is that the PV generation is high during the day when the EVs are usually unavailable, and the n-Grids do not have enough BESS charging capacity to store the generated energy (see Fig. 9). The total PV energy spillage was 2.22 MWh.

### C.    Effects of Repair Time on Disconnected n-Grids

Now, we analyze the impact of the utility outage repair time on the performance of the n-Grids. The obtained load loss and spilled energy of n-Grids as functions of the outage repair time are provided in Fig. 11. As expected, the higher the repair time, the higher the load loss and spilled PV power. The load loss raises significantly per repair time because the stored energy of the n-Grid energy resources cannot meet their demand. An approximately linear relationship is seen in the spilled PV power caused by a shortage in energy storage during the day.

The n-Grid flexibility illustrated in this example allows optimal energy usage supplied by the grid. When that energy is unavailable due to the grid outage, the n-Grid resources continue to supply energy to its load. The n-Grid energy resources are renewable, and hence the deployment of such resources leads to net-zero carbon emissions. At the same time, the bulk power supply can also be substituted with the renewables, further increasing the net-zero grid objectives because the n-Grid flexibility can be an additional resource to manage the overall grid flexibility at times when the bulk renewables experience contingencies due to weather conditions.

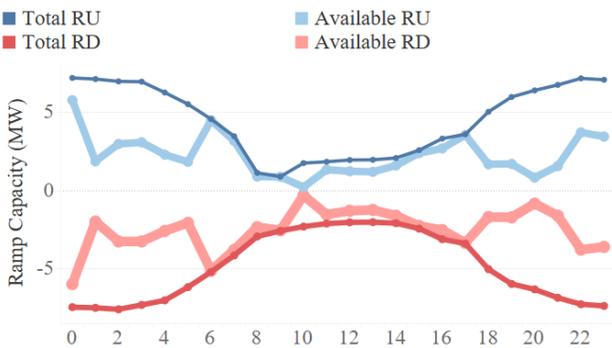

Figure 9. Total RU and RD versus available RU and RD.

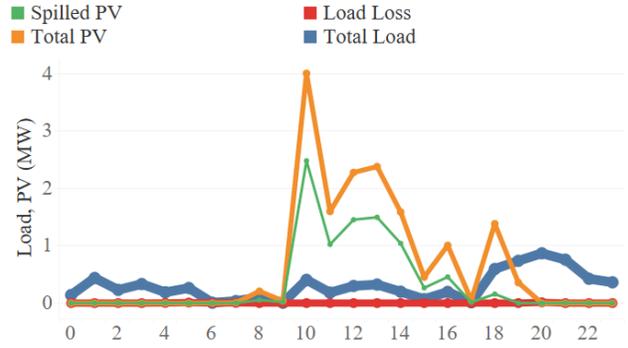

Figure 10. Total load, PV generation, load loss, and spilled PV.

### VI.    CONCLUSIONS

The study we performed points to a few important conclusions:

- The net-zero grid goals benefit from deploying renewables as the bulk energy supply for the utility grid and at the n-Grid level as the customer on-site alternative energy supply for the local load.

- To utilize the flexibility in using the bulk energy supply vs. distributed n-Grid supply, proper planning of the energy resources in the grid is needed at different time horizons, from operations planning to real-time operation and WEM ASP services.

- To offer n-Grid flexibility, the SoR prediction for the utility grid outage and n-Grid demand is needed, which then can be used to optimize, plan and, in some instances, execute the mitigation measures ahead of time.

- The risk prediction approach we described requires extensive use of data-based models incorporating historical data and a variety of weather and other environmental data, all fused through the ML model development.

- The utilities, n-Grid owners, aggregators, and WEM operators may use the outcome of the predictive SoR models to improve their planning and coordination of multiple services n-Grids can offer.

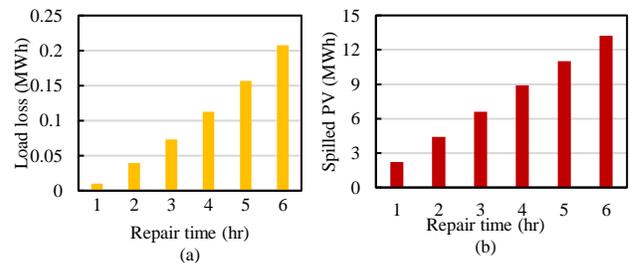

Figure 11. Total load loss and spilled PV as functions of repair time are given in (a) and (b), respectively.

- If the n-Grid owners invest in their energy resources, they need to be able to recover their investment in a reasonable time, so the n-Grid services to support utility and WEM ASP needs should be compensated if they add to the grid resilience.

ACKNOWLEDGMENT

The authors wish to acknowledge the financial support, expertise, and historical outage data provided by CenterPoint Energy and United Cooperative Services through the TEES Smart Grid Center over the years.